\documentclass{article}

\usepackage[utf8]{inputenc} 
\usepackage[T1]{fontenc}    
\usepackage{hyperref}       
\usepackage{url}            
\usepackage{booktabs}       
\usepackage{amsfonts}       
\usepackage{nicefrac}       
\usepackage{microtype}      
\usepackage{amsmath}
\usepackage{amssymb}
\usepackage[pdftex]{graphicx}
\PassOptionsToPackage{numbers,sort&compress}{natbib}
\usepackage[final]{tackling_climate_workshop_style}
\usepackage{xcolor}

\title{HybridCT: ML Accelerated Tomography \\ for Climate Relevant 3D Cloud Observations}
\author{%
  Linda Forster\\
  Jet Propulsion Laboratory\\
  California Institute of Technology\\
  Pasadena, CA, USA.
  \And
  Nicholas J. LaHaye\\
  Jet Propulsion Laboratory\\
  California Institute of Technology\\
  Pasadena, CA, USA.
  \AND
  Marcin J. Kurowski\\
  Jet Propulsion Laboratory\\
  California Institute of Technology\\
  Pasadena, CA, USA.
  \And
  Anna Jungbluth\\
  Imperial College London\\
  London, UK
  \And
  Bernhard Mayer\\
  Ludwig-Maximilians-Universität\\
  Meteorologisches Institut, Fakultät für Physik\\
  Munich, Germany.\\
  \And
  Anthony B. Davis\\
  Jet Propulsion Laboratory\\
  California Institute of Technology\\
  Pasadena, CA, USA.
}

\begin{document}

\maketitle

\begin{abstract}
Low-level clouds are one of the largest sources of uncertainty in climate predictions, partly because we lack detailed three-dimensional (3D) measurements of their internal structure.
Here, we present HybridCT, a cloud tomographic reconstruction scheme accelerated by machine learning (ML) that recovers 3D cloud extinction directly from passive multi-angle observations.
An ML emulator maps radiances from multiple viewing angles to cloud optical thickness (COT), producing path-integrated projections, and a Simultaneous Algebraic Reconstruction Technique (SART) explicitly inverts these projections to retrieve 3D extinction.
Training observations are generated from synthetic 3D cloud fields using radiative transfer simulations for nine viewing angles between ±70$^{\circ}$.
HybridCT reconstructs extinction with RMSE 13.3~km$^{-1}$ in 3.5~s, compared to 22.7~km$^{-1}$ in approx. 30~min for the state-of-the-art physics-based tomography framework AT3D, a speedup exceeding 500$\times$.
HybridCT thus provides a scalable route toward large-scale observations of 3D cloud structure relevant to radiative-budget interpretation and evaluation of high-resolution cloud simulations.
\end{abstract}

\section{Introduction}
Low-level clouds strongly influence Earth's radiation budget and remain one of the largest sources of uncertainty in climate feedbacks and future warming projections \citep{ipcc2021,hartmann1992,nuijens2019,webb2024,sherwood2020}.
Many processes governing cloud evolution and radiative effects are inherently 3D, yet satellite observations involve a fundamental tradeoff:
Vertically resolved measurements provide detailed profiles but only along limited sampling tracks, while passive imagers provide broad spatial and temporal coverage but mostly of column-integrated or cloud-top properties.
This limits our ability to observe cloud depth, internal structure, and organization over large areas.
Recent studies suggest that such 3D information could improve understanding of cloud processes, radiative effects, and the evaluation of high-resolution atmospheric models \citep{kurowski2023, forster2026}. 
Multi-angle tomography can address this observational gap.
Multiple views of the same cloud provide complementary geometric information for reconstructing 3D structure \citep{levis2015,levis2017,levis2020}.
The main obstacle is computational: physics-based tomography requires repeated 3D radiative-transfer calculations that are expensive for realistic cloud scenes \citep{loveridge2023retrieving1,loveridge2023retrieving2}.
This limits the number and size of cloud fields that can be processed, and consequently the use of tomography for large observational datasets.
While machine learning (ML) is increasingly used to derive 3D cloud structure from 2D observations, we distinguish tomography from other 3D reconstruction techniques, like radiance mapping/extrapolation of single view points \citep{bruning2024, 2410.04135, girtsou2025}. Tomography requires multi-angle observations that constrain 3D information geometrically and assumes the cloud field remains static across viewing angles.
Prior works using multi-angle observations include 3DeepCT \citep{Sde-Chen2020-zr} and VIPCT \citep{Ronen2025-nd} for extinction, and NeMF \cite{Kom-Betzer2025-gd} for microphysical properties. However, these approaches do not explicitly constrain geometry, but learn to extrapolate 3D structure through supervised learning with simulated ground truth cloud volumes. For real observations, ground truth volumes are not available. 
ProbCT \citep{ronen2024learned} partially addresses this through a 3D radiative transfer solver that compares real imagery against renderings from predicted volumes. However, this is again limited by the computational constraint of 3D radiative transfer calculations. 
Neural radiance fields \citep{2003.08934} instead constrain volumes via differentiable rendering, but are optimized per scene and require higher angular sampling and spatial resolutions than spaceborne multi-angle instruments currently provide. 
HybridCT addresses these constraints with a hybrid ML and tomographic approach:
\begin{equation}
\mathbf{L}_{1:N}
\xrightarrow{\mathrm{U\text{-}Net}}
\boldsymbol{\tau}_{1:N}
\xrightarrow{\mathrm{SART}}
\boldsymbol{\beta}_{\mathrm{ext}}(x,y,z).
\end{equation}
\begin{figure}[t]
    \centering
    \includegraphics[width=\linewidth,
    trim=0cm 1.2cm 0cm 4cm]{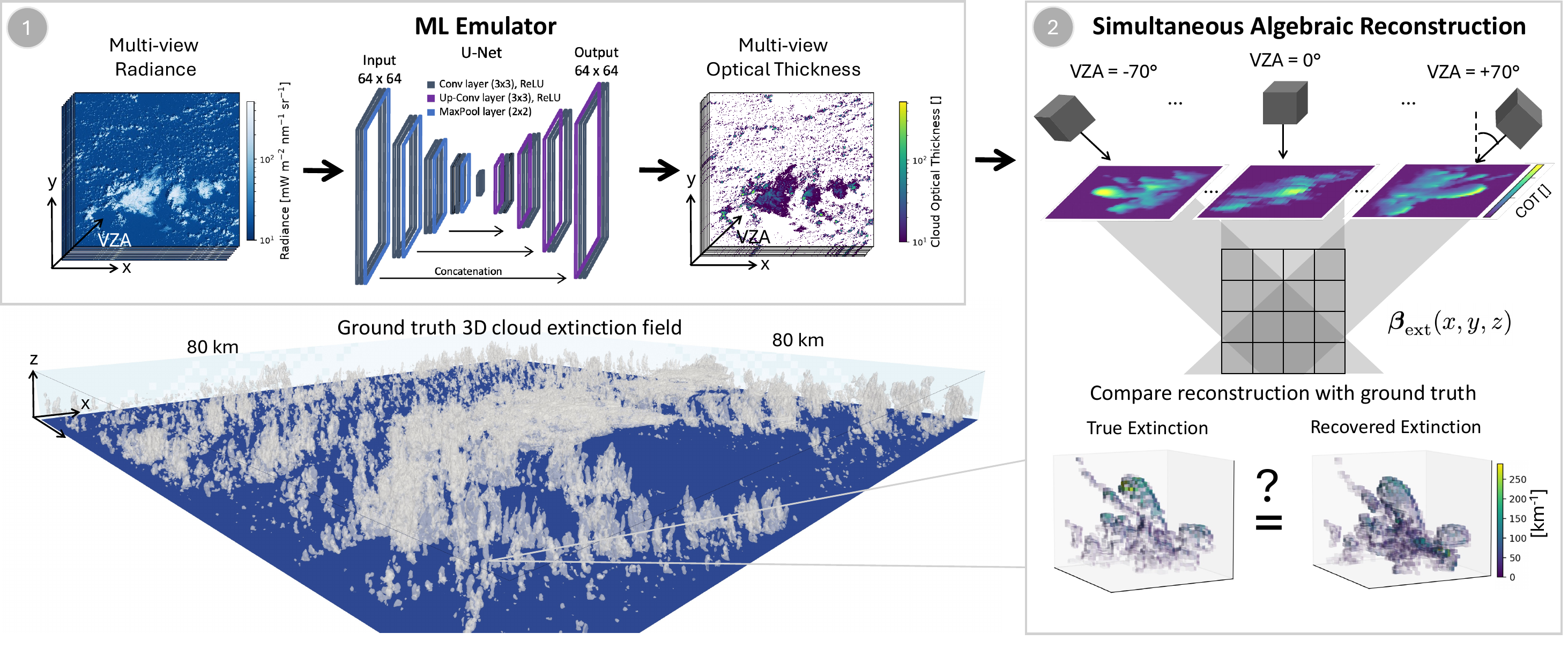}
    \caption{
    Overview of HybridCT.
    (1) Multi-angle passive radiances are first mapped to cloud optical thickness projections using a U-Net.
    (2) The resulting projections are then inverted with an explicit SART tomography solver to reconstruct the three-dimensional cloud extinction field.
    }
    \label{fig:hybridct_overview}
\end{figure}
As illustrated in Figure~\ref{fig:hybridct_overview}, an ML emulator U-Net first converts passive multi-angle radiances $\mathbf{L}$ into cloud optical thickness (COT) $\boldsymbol{\tau}$ projections for $N$ viewing angles.
An explicit algebraic tomography step then reconstructs the 3D extinction field $\boldsymbol{\beta}_{\mathrm{ext}}(x,y,z)$ from those projections.
The neural network therefore solves only the nonlinear radiance-to-COT problem, while the vertical structure remains constrained by the multi-angle observation geometry. 
Our contributions are threefold:
1)~we introduce a hybrid framework in which ML retrieves physically interpretable COT projections and an explicit tomographic solver reconstructs 3D cloud volumes,
2)~vertical structure is constrained by passive multi-angle geometry rather than learned from 3D labels or extrapolated from vertical profiles,
3)~we show that this decomposition reduces reconstruction time by $>500\times$ relative to iterative radiance-space tomography while retaining comparable accuracy.
This provides a step toward scalable, climate-relevant 3D cloud observations that support studies of cloud processes, 3D radiative effects, and evaluation of high-resolution atmospheric models.

\section{Data \& Methodology}

\textbf{Synthetic cloud scene and multi-view radiance.} We demonstrate HybridCT using synthetic observations from large-eddy simulations (LES) performed with the System for Atmospheric Modeling \citep{khairoutdinov2003} for marine subtropical clouds \citep{rauber2007,vanzanten2011}.
Each of the six LES scenes spans $80$~km$\times80$~km at 40~m horizontal resolution and contains realistic 3D fields of shallow cumulus with varying degrees of convective organization and complexity.
Synthetic radiances are generated using the 3D Monte Carlo radiative-transfer solver MYSTIC \citep{mayer2009,buras2010,emde2011} which is part of libRadtran \citep{mayer2005,emde2016}, using 10$^3$ photons per pixel.
The simulated cloud radiance field accounts for a realistic atmospheric background in a maritime environment \citep{anderson86,gasteiger2014,hess98} and ocean-surface reflection assuming a wind speed of 10 m/s \citep{cox54a,cox54b}.
The illumination geometry assumes a sun-synchronous satellite orbit with an equator crossing time in the local AM and the sun in the South-East.
This translates to a solar azimuth angle of 320$^\circ$ and a solar zenith angle (SZA) ranging from 20$^\circ$ to 50$^\circ$.
We simulate the nine viewing directions of the Multi-angle Imaging Spectroradiometer (MISR) \citep{diner1998} between ±70$^{\circ}$ at 670~nm, retaining the 40 m model-grid resolution.
All data used in this work are publicly available on Zenodo \citep{zenodo_2025}.
The HybridCT codebase is available at \url{https://github.com/lforster/hybridCT} and \cite{Forster_HybridCT_2025}.
Version v4.1.2 of AT3D is preserved in \cite{at3d_2022} and is publicly available at \url{https://github.com/CloudTomography/AT3D}.

\begin{figure}[t]
    \centering
    \includegraphics[width=\linewidth, 
    trim=0cm 1.5cm 0cm 4cm]{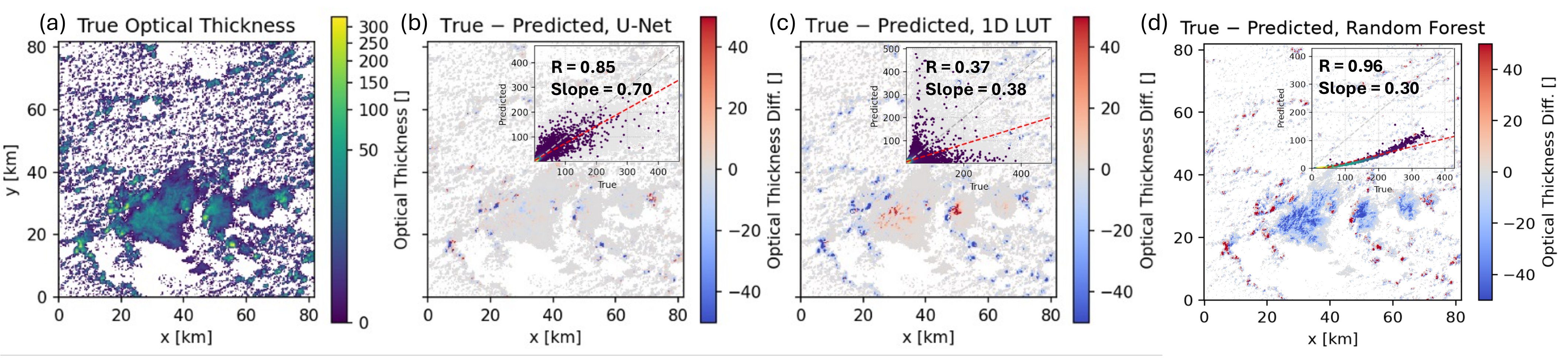}
    \caption{(a) True Cloud optical thickness (COT) field, and true minus predicted COT for (b)  U-Net, (c) 1D look-up table (LUT), and (d) Random Forest regression for the nadir view (VZA = 0°). Scatter plots show correlation coefficient and linear fit slope.
    }
    \label{fig:unet_cot}
\end{figure}

\textbf{Radiance-to-COT retrieval.  } For each viewing angle, a convolutional U-Net \citep{ronneberger2015} emulates the mapping of 2D radiance to COT fields. We train a separate emulator per viewing angle for flexibility when some or different perspectives are available.
Using spatial context and directional information, the network captures 3D radiative effects that 1D look-up tables (LUTs) cannot represent, resulting in higher correlations, reduced bias, and more consistent performance \citep{nataraja2022}.
For training, we split the large cloud and radiance scenes into overlapping square tiles.
Hyperparameter tuning with Python's \emph{ray} module \citep{liaw2018tune} resulted in a 16-pixel tile size, 2-pixel tile stride, learning rate of 0.01, and MinMax scaler.
With a train-test split of 70\% to 30\%, final training over 96,605 samples $\times$ 9 views took about four hours on one GPU with Adam \cite{adam_opt}.
One of the six LES scenes was held out from training and tuning for final evaluation, and all results reported here are based on this held-out scene.
Figure~\ref{fig:unet_cot} compares predicted with true COT for the U-Net (b), 1D LUT (c), and Random Forest (d). Random Forest regression captures the nonlinear radiance-COT relationship but lacks spatial context and remains strongly underestimated at high COT ($R=0.96$ and slope = 0.30).
The 1D LUT, used operationally \citep{nakajima1990, platnick2003}, assumes horizontally homogeneous clouds and neglects 3D radiative effects, resulting in low correlation ($R=0.37$) and strong underestimation (slope $=0.38$) for broken clouds \citep{barker1995}. 
The U-Net provides the best overall agreement, combining high correlation (\(R=0.85\)) with a slope closer to unity (\(0.70\)).
The learned task is deliberately restricted to retrieving 2D path-integrated COT fields, which then serve as projection data for tomography. \newline
\textbf{Multi-view tomography.  } Optical thickness along ray \(i\) is defined as $\boldsymbol{\tau}_i = \int_{\mathrm{LOS},i} \boldsymbol{\beta}_{\mathrm{ext}}(\mathbf r) ds$, with the cloud extinction coefficient $\boldsymbol{\beta}_{\mathrm{ext}}$, the spatial position $\mathbf r$ , and the measurement-ray index \(i\).
After discretization, the forward problem becomes $\boldsymbol{\tau} = \mathbf F \boldsymbol{\beta}_{\mathrm{ext}}$, where $\boldsymbol{\tau}$ is the vector of observed optical thicknesses, $\boldsymbol{\beta}_{\mathrm{ext}}$ is the voxelized 3D extinction field, and $\mathbf F$ represents the multi-angle ray-voxel geometry.
HybridCT solves this inverse problem using the Simultaneous Algebraic Reconstruction Technique (SART) \citep{andersen1984}, iteratively updating $\boldsymbol{\beta}_{\mathrm{ext}}^{(k+1)}=\boldsymbol{\beta}_{\mathrm{ext}}^{(k)}+\lambda\mathbf B[\boldsymbol{\tau}-\mathbf F \boldsymbol{\beta}_{\mathrm{ext}}^{(k)}]$, while enforcing positivity and a prescribed cloud mask, which here is taken from the ground-truth cloud field.
Here, \(k\) denotes the iteration index, $\mathbf F$ and $\mathbf B$ denote the forward-projection and normalized back-projection operators, respectively, and the relaxation parameter $\lambda$ is set to 0.15.
Thus, ML estimates the projection quantities, while the volumetric structure is constrained geometrically rather than learned.

\section{Results \& Conclusions}

\begin{figure}[t]
    \centering
    \includegraphics[width=\linewidth,
    trim=0cm 0.6cm 0cm 4cm]{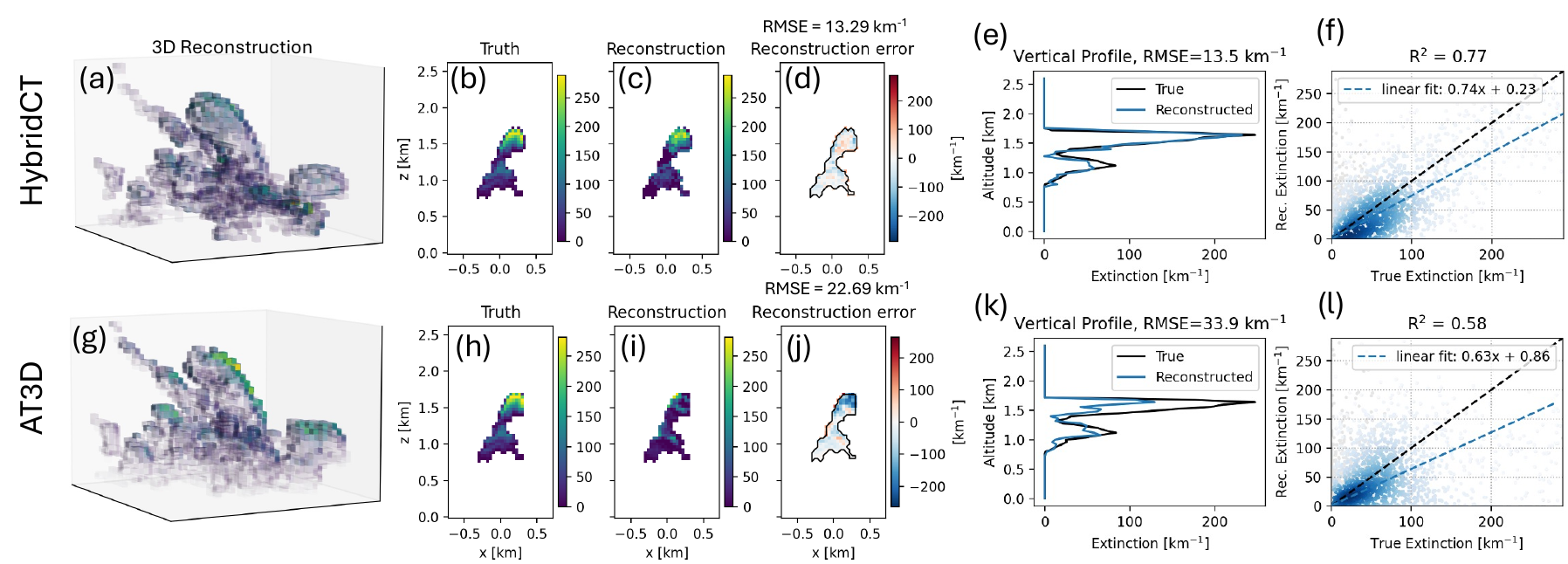}
    \caption{HybridCT (a–f) vs. AT3D (g–l) tomographic reconstruction of cloud extinction.}
    \label{fig:result}
\end{figure}

\begin{figure}[b]
    \centering
    \includegraphics[width=\linewidth,
    trim=0cm 0.7cm 0cm 0.8cm]{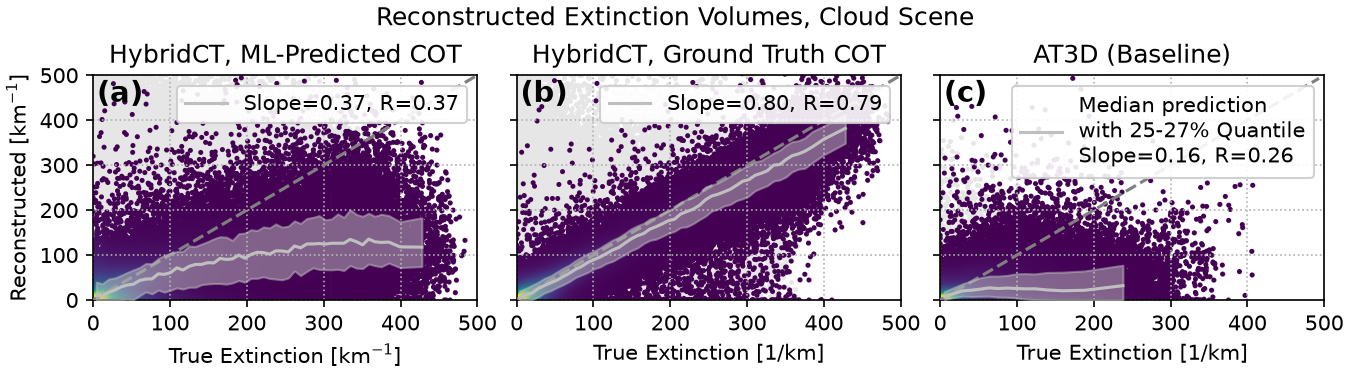}
    \caption{Reconstruction of entire cloud scene showing true vs. reconstructed extinction with correlation coefficient $R$ and slope of the median prediction in each bin (solid gray curve) with 25th–75th percentile range (shaded) for HybridCT with (a) predicted, (b) true COT, and (c) AT3D.}
    \label{fig:result-scatter-plot}
\end{figure}

We first evaluate HybridCT for a shallow cloud approx. 0.9~km across and 1.2~km deep within a larger LES cloud field and compare the result with the physics-based baseline model AT3D (\autoref{fig:result}).
This model reconstructs 3D directly from multi-angle radiances using a 3D radiative-transfer forward operator \citep{evans1998spherical,loveridge2023retrieving1}.
We also tried to compare our results to VIPCT as the closest ML baseline, but their trained checkpoint did not generalize to the viewing and rendering geometries of our data.
HybridCT reconstructs extinction with RMSE $=13.29$~km$^{-1}$, $R=0.77$, a $0.74$ slope, while retaining second-scale runtime.
AT3D reaches RMSE $=22.69$~km$^{-1}$ and $R\approx0.58$ with a $0.63$ slope after ~30~min.
Figure~\ref{fig:result-scatter-plot} compares HybridCT and AT3D over the full cloud field. HybridCT (a) reconstructs more than 6300 clouds in under 6~h with $R=0.40$ and slope $=0.37$, whereas AT3D (c) reconstructs 20 randomly selected clouds in 18~h with lower correlation ($R=0.26$) and slope $=0.16$.
Figure~\ref{fig:result-scatter-plot}b shows results for HybridCT in the theoretical case with error-free COT prediction.
A similar experiment with the cloud in Figure~\ref{fig:result} reduced the RMSE from 13.3 to 6.2~km$^{-1}$, while isolating the target cloud from neighboring clouds further reduced it to 3.76~km$^{-1}$ (not shown here).
This highlights two main error sources for both HybridCT and AT3D: (1) the conversion from radiance to COT and (2) contributions from neighboring clouds that intersect with oblique viewing rays and contaminate individual projections.

In summary, HybridCT differs from fully learned multi-view tomography \citep{sdechen2021,ronen2022variable,ronen2024learned} by breaking the problem into two components:
The neural network predicts only COT projections, while the 3D extinction field must satisfy an explicit multi-view geometric projection operator.
This restricted role for machine learning makes the source of vertical information physically transparent, and ensures ML is only used where it is most useful. Moreover, since HybridCT’s tomographic reconstruction is geometrically constrained rather than learned, it transfers directly to other datasets (e.g., the LES clouds from VIPCT \citep{ronen2022variable}) without retraining or distribution-shift issues that might arise from applying models to unseen data.
The climate relevance follows two main pathways.
First, cloud geometry influences how radiation propagates through and around heterogeneous cloud fields.
Observations and 3D radiative-transfer studies show that neglecting this structure can produce substantial radiative errors, particularly for broken shallow clouds \citep{gristey2020surface,villefranque2021,singer2021}.
Retrieved 3D extinction and geometry therefore provide direct observational constraints for studying and evaluating these effects.
Second, high-resolution atmospheric models increasingly resolve cloud structures and processes that remain parameterized in coarser models, creating a need for observations against which those simulations are evaluated \citep{schneider2017climate,schneider2017esm}.

While multi-angle cloud tomography offers a route toward volumetric cloud observations, the computational cost of repeated 3D radiative-transfer calculations has so far prevented its application to large, realistic cloud scenes \citep{levis2015,levis2017,loveridge2023retrieving1,loveridge2023retrieving2}.
HybridCT directly targets this computational barrier by moving the expensive radiative inversion into a learned preprocessing step while retaining an explicit geometric reconstruction. The latter runs easily on the CPU of a personal computer and is orders of magnitude faster than comparable physics-based models.
Important limitations remain.
The present study uses synthetic observations from one cloud regime, the U-Net is trained for MISR-like viewing geometry and a limited range of environmental conditions, and COT does not preserve all information contained in the original radiances.
Real-data application will require broader training, observation-based 3D cloud masking, and independent validation with active and in situ observations.
The static-scene assumption generally requires near-simultaneous multi-angle acquisition which will be achieved by next-generation satellite constellations such as CloudCT \citep{schilling2019cloudct}.
HybridCT demonstrates that learned radiance inversion can be combined with explicit passive tomography to reduce reconstruction from hours to seconds while retaining useful 3D fidelity.
This opens a pathway toward volumetric retrievals at regional to global scales, providing observations of internal extinction structure and cloud organization for studying 3D radiative effects and evaluating cloud-resolving and climate models.

\begin{ack}
The research was carried out at the Jet Propulsion Laboratory,
California Institute of Technology, under a contract with the National
Aeronautics and Space Administration (80NM0018D0004).
The High Performance Computing resources used in this investigation were provided by funding from the JPL Information and Technology Solutions Directorate.
The authors acknowledge the Texas Advanced Computing Center (TACC) at The University of Texas at Austin for providing computational resources that have contributed to the research results reported within this paper \url{https://www.tacc.utexas.edu}.
We thank Aviad Levis, Katherine L. Bouman, and Steffen Mauceri for fruitful discussions.
\end{ack}

\small
\bibliographystyle{unsrt}
\bibliography{literature}


\end{document}